\documentclass[prb,aps]{revtex4}
\usepackage{graphicx}
\usepackage{amsmath,amssymb}

\begin{document}

\title{Many-body treatment of quantum transport through single molecules}


\author{J.~P.~Bergfield}
\affiliation{College of Optical Sciences, University of Arizona, 1630 East University Boulevard, AZ 85721}
\author{C.~A.~Stafford}
\affiliation{Department of Physics, University of Arizona, 1118 East Fourth Street, Tucson, AZ 85721}

\begin{abstract}
We develop a 
theoretical approach to quantum transport through single conjugated organic molecules that 
accurately accounts for all molecular excited and ground states via a semi-empirical many-body molecular Hamiltonian. 
We then calculate the linear and non-linear transport properties of a benzenedithiol molecule covalently bonded to two gold electrodes at room temperature, 
eliminating all free parameters by comparing the calculated thermopower and linear-conductance with experiment.  
The resulting `molecular diamond' structure exhibits experimentally observed many-body effects inaccessible to mean-field approaches.
\end{abstract}

\maketitle

Electron transport through single molecules is of interest both fundamentally and as a means to probe numerous important chemical\cite{BioBook99} and biological processes\cite{Schrieber05}, with myriad potential device applications\cite{Nitzan03}.  In spite of the growing body of experimental evidence indicating that intra-molecular interactions play an important role in transport problems involving single-molecule 
junctions\cite{Park00,Liang02,Kubatkin03,Poot06}, 
theoretical studies of these junctions typically focus on either many-body phenomena, such as the Kondo effect\cite{Glazman04}, while neglecting the internal structure of the molecule, or strive for a realistic description of the molecule's electronic structure via 
{\em ab initio} methods\cite{Nitzan03,Lindsay07}, which treat interactions at the mean-field level.  

In this article, we develop a many-body non-equilibrium Green's function (NEGF) approach to the problem of quantum transport in multi-terminal single-molecule devices.  The electronic degrees of freedom most relevant for transport, i.e.,
the $\pi$-electrons in conjugated organic molecules, are modeled via 
a semi-empirical Pariser-Parr-Pople (PPP) Hamiltonian\cite{Ohno64,Chandross97,Castleton02}.  
The molecular Hamiltonian is solved via exact diagonalization, 
thereby including intra-molecular correlations non-perturbatively.  
The coupling of the molecule to the macroscopic leads is treated perturbatively within the NEGF formalism\cite{Meir92,Jauho94,Datta95}.

Electronic transport is a property of a junction: molecule plus leads.  The Hamiltonian of a junction may be written as 
$ H=H_{\rm mol}+H_{\rm leads}+H_{\rm tun}$, where the first term is the $\pi$-electron molecular Hamiltonian\cite{Cardamone06}
\begin{equation}
\label{eq:H_mol}
 H_{\rm mol}=\sum_{n\sigma}\varepsilon_n d_{n\sigma}^\dagger d_{n\sigma}
-\sum_{\rm \langle nm\rangle\sigma}\left(t_{nm}d_{n\sigma}^\dagger
  d_{m\sigma}+\mathrm{H.c.}\right)
\mbox{ }+ \sum_{nm}\frac{U_{nm}}{2}Q_nQ_m,
\end{equation}
where $d^\dagger_{n\sigma}$ creates an electron of spin $\sigma$ in the $\pi$-orbital of the $n$th carbon atom, $\varepsilon_n$ is the on-site orbital energy, $t_{nm}$ are the tight-binding matrix elements between orbitals $n$ and $m$,  $U_{nm}$ are the intra- and intersite Coulomb interactions, and $\scriptstyle \langle \mbox{ } \rangle$ indicates a sum over nearest-neighbors.

The effective charge operator\cite{Stafford98} for orbital $n$ is given by
\begin{equation}
Q_n=\sum_\sigma d_{n\sigma}^\dagger d_{n\sigma}-\sum_\alpha C_{n\alpha}\mathcal{V}_\alpha/e -1,
\label{eq:Qn}
\end{equation}
where
$C_{n\alpha}$ is the capacitive coupling between orbital $n$ and lead $\alpha$, $e$ is the electron charge, and $\mathcal{V}_\alpha$ is the voltage on lead $\alpha$.  The lead-molecule capacitance values are elements of a full capacitance matrix,\cite{CapMatrixBook84} $C^{\rm full}$, which includes both the intra-molecular and inter-molecular capacitances.  This full capacitance matrix obeys the usual zero-sum rules required for gauge invariance and is constructed in accordance with the intra-molecular interaction energies (i.e. $C^{\rm full}_{nm}=e^2U^{-1}_{nm}$) and the geometry of the junction, where we make the ansatz that $C_{n\alpha}$ is inversely proportional to the lead-site distance. 

The two remaining terms in the junction's Hamiltonian, $H_{\rm leads}$ and $H_{\rm tun}$, account for the isolated leads and 
lead-molecule coupling, respectively.  Due to their large density of states, the macroscopic metallic leads (labeled $\alpha\in[1,\ldots,M]$) may be modeled as non-interacting Fermi gases, with tunneling matrix elements $V_{nk}$ from a level $k$ of energy $\epsilon_k$ within lead $\alpha$ to a nearby $\pi$-orbital $n$ of the molecule.\cite{Cardamone06}  These terms are given explicitly by
\begin{equation}
H_{\rm leads}=\sum_{\alpha=1}^M\sum_{\substack{k\in\alpha\\ \sigma}}\epsilon_k c_{k\sigma}^\dagger c_{k\sigma}
\label{eq:H_leads}
\end{equation}
and
\begin{equation}
H_{\rm tun}=\sum_{\langle n\alpha\rangle}\sum_{\substack{k\in\alpha\\ \sigma}}\left(V_{nk}d_{n\sigma}^\dagger
  c_{k\sigma}+\mathrm{H.c.}\right),
\label{eq:H_tun}
\end{equation}
where $c^\dagger_{k\sigma}$ is a lead-electron creation operator.

The effective interaction energies for a $\pi$-conjugated polymer are given by \cite{Ohno64,Chandross97,Castleton02}
\begin{equation}\label{eq:ohno}
U_{nm}=\delta_{nm}U+\left(1-\delta_{nm}\right)\frac{U}{\epsilon  \sqrt{1+\alpha(R_{nm}/\mbox{\AA})^2}},
\end{equation}
where $U$ is the on-site Coulomb energy, $\epsilon$ is the dielectric constant, $\alpha=\left( U/14.397\mbox{eV} \right)^2$  
, and $R_{nm}$ is the distance between orbitals $n$ and $m$.  The dielectric constant $\epsilon$ accounts for screening 
effects due to both the $\sigma$-electrons and any environmental considerations, such as any non-evaporated solvent.\cite{Castleton02}  The solutions of the PPP Hamiltonian, used in conjunction with these experimentally derived interaction energies, accurately account for all ground and excited state energies and wavefunctions of the free molecule\cite{Castleton02}.

When the isolated molecule, with its discrete molecular energy spectrum, is coupled to macroscopic metallic leads a 
junction is formed and the bare molecular energy levels become dressed by interactions with the metal's electronic 
state continuum.  These interactions give rise to width broadened and shifted molecular energy levels which are imbued, correspondingly, with finite lifetimes. We are interested in the non-equilibrium steady state behavior of these junctions both as a function of bias voltage and of molecular gating voltage. 

A considerable simplification to the problem of quantum transport in this fundamentally open system is obtained through the use of the Green's functions $G(E)$ and $G^<(E)$, which are Fourier transforms of the retarded and Keldysh Green's functions,
$G_{n\sigma,m\sigma'}(t)=-i\theta(t)\langle \{d_{n\sigma}(t),d_{m\sigma'}^\dagger(0)\}\rangle$ and
$G^<_{n\sigma,m\sigma'}(t)=i\langle d_{m\sigma'}^\dagger(0)\, d_{n\sigma}(t) \rangle$, respectively\cite{Meir92,Jauho94,Datta95}.  
The intuitive interpretation of the retarded Green's function as the total probability amplitude for a system with a particle (electron or hole) 
at site $n$ to be found a time $t$ later in a state with an added particle at site $m$, is its principle advantage in solving quantum transport 
problems.  
For an isolated molecule, we derive
the (retarded) Green's function from the molecular Hamiltonian and find
%
\begin{equation}
\left[G_{\rm mol}(E)\right]_{n\sigma,m\sigma'} =
\sum_\nu {\cal P}(\nu) \left[
\sum_{\nu'} \frac{\langle \nu | d_{n\sigma} |\nu'\rangle \langle\nu'| d_{m\sigma'}^\dagger |\nu\rangle}{E - E_{\nu'} + E_\nu + i 0^+} +
\sum_{\nu'} \frac{\langle \nu | d_{m\sigma'}^\dagger |\nu'\rangle \langle\nu'| d_{n\sigma} |\nu\rangle}{E - E_\nu + E_{\nu'} + i 0^+}
\right],
\label{eq:Gintra}
\end{equation}
%
where $|\nu\rangle$ and $|\nu'\rangle$ are many-body eigenstates satisfying $H_{\rm mol} |\nu\rangle = E_\nu |\nu\rangle$, and ${\cal P}(\nu)$ is the probability that the isolated molecule is in state $|\nu\rangle$.  In linear response, the probabilities ${\cal P}(\nu)$ in Eq.\ (\ref{eq:Gintra}) are determined by equilibrium thermodynamics,
${\cal P}(\nu) = e^{-\beta(E_\nu - \mu N_\nu)}/{\cal Z}$,
where ${\cal Z}$ is the grand partition function of the molecule at inverse temperature $\beta$ and chemical potential $\mu$.
For nonlinear transport, ${\cal P}(\nu)$ is given by the appropriate limit as $V_{nk} \rightarrow 0$. 
This is the limit of {\em sequential tunneling}, where ${\cal P}(\nu)$ can be determined from detailed balance in a rate-equation approach \cite{Beenakker91}.  

The Green's functions of the full system are
\begin{equation}
\label{eq:Keldysh}
G=\left(G_{\rm mol}^{-1}-\Sigma\right)^{-1}, 
\;\;\;\;\;\; 
G^< = G \, \Sigma^< G^\dagger, 
\end{equation}
%
where 
$\Sigma$ and $\Sigma^<$ are the self-energies describing the dressing of the bare molecular states by the lead-molecule coupling.  Given the Green's functions, the current flowing into lead $\alpha$ is then \cite{Meir92} 
\begin{equation}
I_\alpha = -\frac{e}{h} \int dE\; \Gamma_\alpha(E) \, \mbox{Im} [G^<_{a\sigma,a\sigma}(E) + 2 f_\alpha(E) G_{a\sigma,a\sigma}(E)],
\label{eq:Meir_Wingreen}
\end{equation}
where $\Gamma_\alpha(E)=2\pi\sum_{k\in\alpha}|V_{nk}|^2\delta\left(E-\epsilon_{ k}\right)$ is the tunneling width for lead $\alpha$, $f_\alpha(E)=\{1+\exp[(E-\mu_\alpha)/k_B T]\}^{-1}$, and $a$ is the $\pi$-orbital coupled to lead $\alpha$.

It is nontrivial to derive exact expressions for the self-energies describing lead-molecule coupling.
Wick's theorem cannot be used in the usual way to derive perturbative corrections because $H_{\rm mol}$ is not quadratic in fermion operators.
Instead, diagrammatic expansions must be done in the independent-particle basis, where the rightmost term in Eq.\ (\ref{eq:H_mol}) is also treated
as a perturbation.
To lowest (zeroth) order in $U_{nm}$, the self-energies are simply 
\begin{equation}
\label{eq:Sigma}
\Sigma_{n\sigma,m\sigma'}(E)=\delta_{nm}\delta_{\sigma\sigma'}\sum_{\langle a\alpha\rangle} \delta_{na}
\sum_{k \in \alpha} \frac{|V_{nk}|^2}{E - \epsilon_k + i 0^+},
\end{equation}
and
\begin{equation} 
\label{eq:SigmaLess}
\Sigma^<_{n\sigma,m\sigma'}(E)= i \delta_{nm}\delta_{\sigma\sigma'}\sum_{\langle a\alpha\rangle}f_\alpha(E)\Gamma_\alpha(E)\delta_{na}.
\end{equation}
Even in this lowest-order expression for the self-energy, the lead-molecule coupling is still treated at the same level of
approximation as in mean-field {\em ab initio} models\cite{Nitzan03,Lindsay07}, 
where lead-molecule correlations are neglected. Eqs.\ (\ref{eq:Sigma}-\ref{eq:SigmaLess}) represent a {\em conserving approximation}; 
current is conserved and the spectral function obeys the usual sum--rule. Transport is treated at the level of elastic cotunneling
theory\cite{Averin90}, while sequential tunneling results \cite{Beenakker91} are recovered in the limit $\Gamma_\alpha / T \rightarrow 0$.  This approximation is valid when there are no unpaired electrons in the molecule or at temperatures large compared with the Kondo temperature.\cite{Glazman04}

Within this approximation, we are left with a key result: $G_{\rm mol}$ alone encapsulates the intra-molecular correlations.  
The full NEGF current expression (\ref{eq:Meir_Wingreen}) then reduces to the well-known multi-terminal B\"uttiker formula\cite{Buttiker86}
\begin{equation}
\label{eq:Buttiker}
I_\alpha=\frac{2e}{h}\sum_{\beta=1}^M\int_{-\infty}^\infty
dE\;T_{\beta\alpha}(E)\left[f_\beta(E)-f_\alpha(E)\right],
\end{equation}
where\cite{Datta95} 
\begin{equation} \label{eq:Transmission}
T_{\beta\alpha}(E)=\Gamma_\beta(E)\Gamma_\alpha(E)\left|\left[G(E)\right]_{ba}\right|^2
\end{equation}
is the transmission probability from lead $\alpha$ to lead $\beta$ and $a\, (b)$ is the orbital coupled to lead $\alpha\, (\beta)$. 

Given the large uncertainties concerning the lead-molecule coupling \cite{Baranger05} we shall
take the so-called broad-band limit \cite{Jauho94} in which the $\Gamma_\alpha$ are taken as constants characterizing
the lead-molecule coupling.  Typical estimates \cite{Tian98,Nitzan01,Muralidharan06} using the method of Ref.~26 yield $\Gamma_\alpha \lesssim 1\mbox{eV}$ for organic molecules coupled to gold contacts via thiol groups.  
In the broad-band limit, we find that the non-equilibrium weighting function becomes
${\cal P}\left(\nu\right) =\sum_{\alpha = 1}^{M} \Gamma_\alpha {\cal P}_\alpha \left(\nu\right) / \sum_{\alpha=1}^M \Gamma_\alpha$, 
where ${\cal P}_\alpha \left(\nu\right)$ is the equilibrium probability for a state $\nu$ in the grand-canonical ensemble at the temperature and 
chemical potential of lead $\alpha$.

As a first application of our approach, we consider the well studied two-terminal 1,4-benzenedithiol (BDT) molecular junction with gold leads.\cite{Toher07,Kriplani06,Xiao04, Dadosh05,Reed97}  Systematic molecular energy-level studies of experimental condensed-phase benzene spectra have found the parameterization 
$t_{nm}=2.64\mbox{eV}$, $U=8.9\mbox{eV}$, and $\epsilon=1.28$ of Eqs.\ (\ref{eq:H_mol}) and (\ref{eq:ohno}) 
to give the best quantitative agreement for all ground and excited states.\cite{Castleton02}  Simulations discussed in this paper were all performed with symmetric capacitive and lead-molecule couplings for a junction at room temperature (T=273K). 

The transmission probability given in Eq.\ (\ref{eq:Transmission}) is also the linear-response conductance in units of 
the conductance quantum $G_0=2e^2/h$.  This conductance is a function of both the lead-molecule tunneling width, $\Gamma$, 
and the difference between the lead's Fermi level and the isolated molecule's HOMO--LUMO gap center, $\mu_0$.  Our model therefore has two phenomenological parameters which we shall determine by fitting to experiment.

\begin{figure}[htb]
\centering
\includegraphics[width=3in]{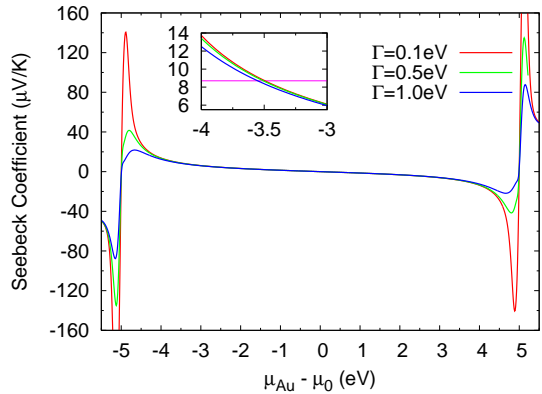} 

\caption[BDT Thermopower]{The Seebeck coefficient as a function of the lead-molecule chemical potential mismatch for BDT-junction. Inset: A closer look at the curve showing, away from a resonance, that the Seebeck coefficient is only weakly dependent on the lead-molecule coupling $\Gamma$. The horizontal line represents the mean experimental value of $8.7 \mu V/K$.\cite{Reddy07}  We find that $\mu_{\rm Au}$ is 3.5eV below the center of the LUMO--HOMO gap ($\mu_0$) of BDT}
\label{fig:BDT_Seebeck}
\end{figure}

Thermopower (Seebeck coefficient) measurements\cite{Reddy07} offer an experimentally accessible
means to determine\cite{Paulsson03b} the lead-molecule chemical potential mismatch, $\mu_{\rm Au}-\mu_0$,  
where $\mu_0= \left(\epsilon_{HOMO}+\epsilon_{LUMO}\right)/2$ and $\epsilon_{HOMO}$ ($\epsilon_{LUMO}$) represents the HOMO (LUMO) energy level.  Although usually associated with bulk materials, the Seebeck coefficient, $S$,  can also be obtained for a molecular junction by measuring the voltage difference created across the junction in response to a temperature differential.
Beginning from Eq.\ (\ref{eq:Buttiker}) and the definition of thermopower as $\Delta {\cal V} = -S\Delta T$,
where ${\cal V}$ is the voltage and T is the temperature, one finds\cite{vanHouten92}
\begin{equation}
\label{eq:Seebeck_FullExpression}
S(\mu)=-\frac{1}{\mbox{eT}}\frac{ \int_{-\infty}^\infty  T_{\beta\alpha}(E) \left( -\frac{\partial f}{\partial E} \right) \left( E-\mu \right) dE}{\int_{-\infty}^\infty  T_{\beta\alpha}(E) \left( -\frac{\partial f}{\partial E} \right)dE },
\end{equation}
where $\mu=\mu_\alpha=\mu_\beta$ is the chemical potential of both leads.  An important feature of the Seebeck coefficient is that asymmetries in conductance with respect to $\mu_{\rm Au}-\mu_0$ manifest themselves in the sign and magnitude of the above integral, thereby lending insight into the nature of the charge transport in an energy barrier dependent, but molecule number independent manner.\cite{Paulsson03b,Reddy07}
Reddy et al.\cite{Reddy07} have measured the Seebeck coefficient of a BDT-junction and found it to be $(8.7\pm2.1)\mu V/K$.
By equating this experimental value with that calculated using Eq.\ (\ref{eq:Seebeck_FullExpression}) we find, as shown in Figure~\ref{fig:BDT_Seebeck}, that $\mu_{\rm Au}-\mu_0=-$(3.5$\pm$0.25)eV.  This value indicates that the Fermi level of the gold lies about $1.5\mbox{eV}$ above the HOMO resonance, which validates the notion that these devices are dominated by p-type (hole) transport.

The only other free parameter is the lead-molecule coupling, $\Gamma$, which can be found by matching 
the linear-response conductance to experiment.  Although there is a large range of experimental values \cite{Lindsay07} the most reproducible and lowest resistance contacts were obtained by Xiao, Xu and Tao~\cite{Xiao04}, who report a single-molecule conductance value of 0.011G$_0$.  This fixes $\Gamma$=(0.53$\pm$0.11)eV, within the range predicted by other groups for similar molecules.  With the final parameter fixed, 
we can now use our many-body approach to predict the linear and nonlinear response of this molecular junction.

\begin{figure*}[htb]
\centering
\includegraphics[width=7in]{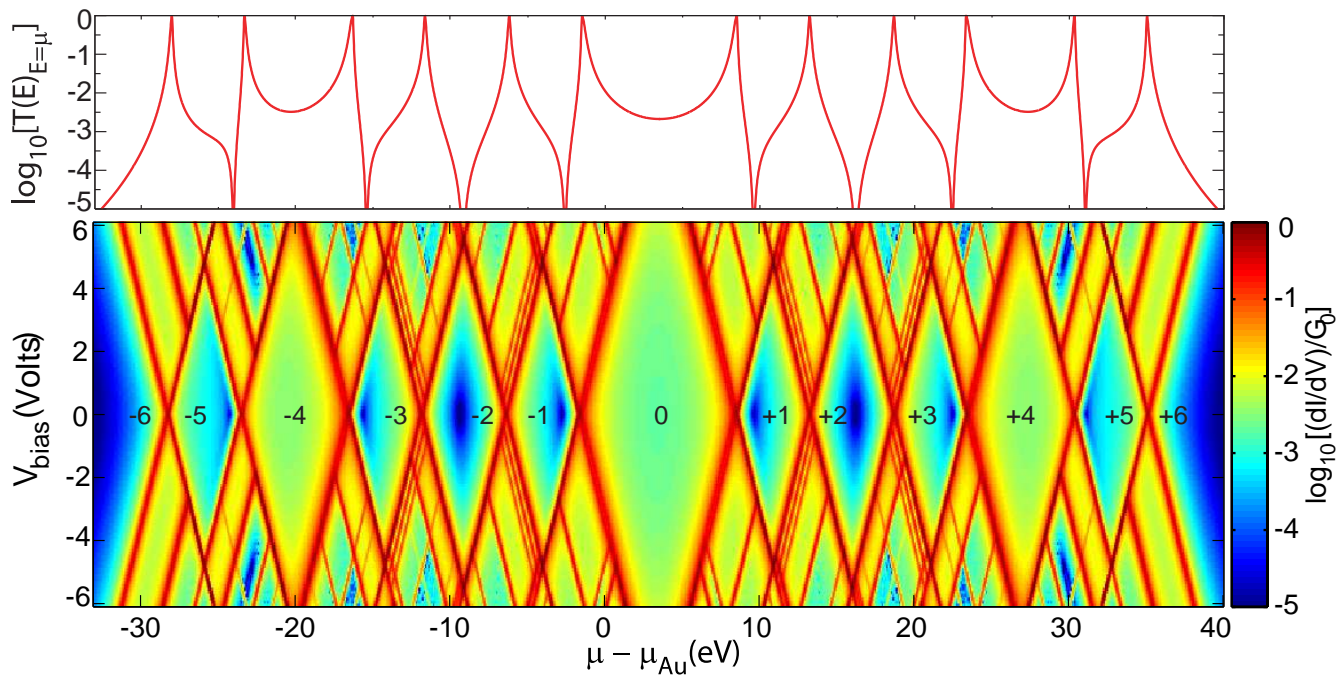}  
\caption[Molecular Diamond]{The linear and non-linear response of a BDT-junction with two gold leads ($\mu_{\rm Au}-\mu_0=-$3.5eV).  Top: The electronic addition spectrum as a function of gating potential at zero bias exhibits non-symmetric, Fano-like resonance lineshapes and a widened LUMO--HOMO gap due to Mott-Hubbard correlations.  This is also the linear-response conductance in units of the conductance quantum $G_0=2e^2/h$.  Bottom: Each column is the logarithm of the differential conductance, $\log_{10}\left[\left(dI/dV\right)/G_0\right]$, calculated for a given gating potential $\mu-\mu_{\rm Au}$.  
The number placed inside each constant-particle-number diamond represents the molecule's charge state, with respect to neutral benzene.  The conductance resonance lines with positive slope indicate hole-like excitations while the negative slope indicate particle-like transport, which at zero bias are degenerate.  The finer, nearly parallel lines inside the outline of the particle-like and hole-like resonances arise from excited state transport.}
\label{fig:FullCoulombDiamond}
\end{figure*}

Initially, a calculation of the transmission probability $T(E)|_{E=\mu}$ as a function of $\mu-\mu_{\rm Au}$ was performed for the BDT-junction by exact diagonalization of Eq.\ (\ref{eq:H_mol}) and the construction of the requisite Green's functions.  The chemical potential $\mu$ is related to the junction's gate voltage $V_g$ via the capacitance $C_g$.  The resulting electron addition spectrum, shown in the top part of Figure~\ref{fig:FullCoulombDiamond}, exhibits several striking features arising from our many-body treatment: an irregular Coulomb blockade-like peak spacing, indicative of charge quantization; an increased LUMO--HOMO gap, reflecting incipient Mott-Hubbard correlations; and non-symmetric, Fano-like resonance lineshape arising from the non-zero thermodynamic weighting for multiple states near a resonance.  Further, since the thermal energy $k_B T$ is far less than any molecular electronic resonance or $\Gamma$, the zero bias addition spectrum is also a map of the ground state electronic structure of BDT.  Our approach therefore reproduces the key results of both the Coulomb blockade and resonant tunneling regimes simultaneously, including intramolecular
many-body correlations exactly.

We next calculate the differential conductance, $\partial I / \partial V_{\rm bias}$, as a function of $\mu$ and $V_{\rm bias}$.  The results of this calculation, shown in the lower part of Figure~\ref{fig:FullCoulombDiamond}, exhibit a `molecular diamond' structure which is a molecular analog of the Coulomb-diamond measured in many quantum dot transport experiments.\cite{DeFranceschi01}  Apart from the central LUMO--HOMO gap, the resonance spacing in the top half of Figure~\ref{fig:FullCoulombDiamond} can be roughly explained via a capacitive model in which the molecule is characterized by a single capacitance $C=e^2/\left\langle U_{nm}\right\rangle$, where $\langle \rangle$ indicates an average over site number.  With the aforementioned parameterization\cite{Castleton02} $\left\langle U_{nm}\right\rangle \sim$5.11eV whereas the LUMO--HOMO gap $\sim$10eV, indicating significant deviations from a simple capacitive charging model due to molecular structure (note also the enlarged diamonds with charge $\pm 4$).  While charge quantization has not yet been observed in BDT junctions due to the difficulty of gating such small molecules\cite{Li06}, the unambiguous observation of Coulomb blockade in junctions involving larger molecules\cite{Kubatkin03,Liang02,Poot06,Danilov08,Park00}, with {\em smaller} charging energies, indicates that such interaction effects, lying outside the scope of mean-field approaches, are undoubtedly even more important in small molecules, like benzene.

A cross section at zero bias reproduces the addition spectrum shown in the top part of the same figure.  Increasing the bias voltage, 
we find that the resonances split into negatively sloped particle-like ($|e|dV/d\mu = -\left(C_1+C_2+C_g\right)/C_1$) lines and positively 
sloped hole-like ($|e|dV/d\mu = +\left(C_1+C_2+C_g\right)/C_2$) lines, where C$_1$ and C$_2$ are the mean lead-molecule capacitances.  In the symmetric coupling case ($C_1=C_2$ with $C_g/C_1\ll 1$) the lines therefore have slopes of -2 and +2 for particle-like and hole-like lines, respectively.

Within the V-shaped outline traced by the particle-like and hole-like lines we find evidence for excited state transport in the many narrow, nearly parallel resonance lines. The structure of these resonances is highly dependent on particle number, as one would expect from the underlying many-body symmetries.  Excited state transport has been observed in quantum dot\cite{DeFranceschi01} and nanotube experiments\cite{Sapmaz06} but hasn't yet been unambiguously identified in single-molecule junctions.

In conclusion, our theoretical approach combining a semi-empirical PPP Hamiltonian with NEGF methods provides
a computationally tractable means to study intra-molecular many-body correlations and excited state transport in a quantitatively accurate way. Applying our theoretical approach to a BDT-junction with two gold leads, we solved for the only two free parameters in our model, finding $\mu_{\rm Au}-\mu_0=-(3.5\pm0.25)\mbox{eV}$ and $\Gamma=(0.53\pm0.11)\mbox{eV}$, indicating that the junction's charge transport is dominated by holes 
(p-type transport) and that the lead-molecule coupling is within the expected range.  
At non-zero bias, the differential conductance of a BDT junction is predicted to exhibit a `molecular diamond' structure analogous to the `Coulomb diamonds' observed in quantum dots\cite{DeFranceschi01},
but with significant modifications due to molecular structure.

The authors acknowledge useful discussions with S.\ Mazumdar and D.\ M.\ Cardamone.

\providecommand{\url}[1]{\texttt{#1}}
\providecommand{\refin}[1]{\\ \textbf{Referenced in:} #1}

\end{document}